\documentclass[aps,prb,twocolumn,groupedaddress,showpacs]{revtex4-1}
\usepackage[bookmarks]{hyperref}
\usepackage{graphicx}
\usepackage{amsmath}
\usepackage{amstext}
\usepackage{graphics}
\usepackage{exscale}
\usepackage{epsfig}
\usepackage[T1]{fontenc}
\usepackage{bm}
\usepackage{bbm}

\usepackage{version}

\usepackage{latexsym}

\usepackage[tight]{subfigure}

\renewcommand{\epsilon}{\varepsilon}

\newcommand{\bra}[1]{ \langle \! #1\ \! |} 
\newcommand{\ket}[1]{\! | #1\ \!\! \rangle}

\newcommand{\integral}[3]{\!\int\limits_{#2}^{#3}\!\!{\rm d}#1\;}

\newcommand{\elcre}[2]{ c^{\dagger}_{#1,#2}}
\newcommand{\elann}[2]{ c^{}_{#1,#2}}

\newcommand{\e}{\mathrm e}

\newcommand{\vk}{{\bm k}}

\newcommand{\hc}{\mathrm{h.c.}}

\begin{document}

\title{Analysis of low energy response and possible emergent SU(4) Kondo
state in a double quantum dot}
\author{Yunori Nishikawa,${}^{1,2}$  Alex C. Hewson,${}^2$ Daniel J.G. Crow,${}^2$
  and Johannes Bauer${}^{3}$}
\affiliation{${}^1$Department of Physics, Osaka City University, Sumiyoshi-ku,
Osaka 558-8585 Japan}
\affiliation{${}^2$Department of Mathematics, Imperial College, London SW7 2AZ, UK}
\affiliation{${}^3$Department of Physics, Harvard University, Cambridge,
  Massachusetts 02138, USA}
\date{\today} 
\begin{abstract}
We examine the low energy behavior of a double quantum dot in a regime where
spin and pseudospin excitations are degenerate. The individual quantum dots
are described by Anderson impurity models with an on-site interaction $U$ 
which are capacitively coupled by an interdot interaction $U_{12}<U$. The
low energy response functions are expressed in terms of renormalized
parameters, which can be deduced from an analysis of the fixed point in a
numerical renormalization group calculation.  At the point
where the spin and pseudospin degrees of freedom become degenerate, the free
quasiparticle excitations have a phase shift of $\pi/4$ and a 4-fold
degeneracy. We find, however, when the quasiparticle interactions are
included, that the low energy effective model has SU(4) symmetry only in the
special case $U_{12}=U$ unless both $U$ and $U_{12}$ are greater than $D$, the half-bandwidth of the
conduction electron bath. We show
that the gate voltage dependence of the temperature dependent differential
conductance observed in recent experiments can be described by a quasiparticle
density of states  with temperature dependent renormalized parameters. 

\end{abstract}

\maketitle

\section{Introduction}
There has been much recent  theoretical and experimental interest in the low energy
behavior of coupled quantum dots where the electrons are strongly confined
on the dots.\cite{WFEFTK02,PSS01,BZHHD03,GLK05,MGL06,CSGSE09,CC09,OSM11,TVLBCAR12} 
As a consequence of this confinement,  the on-site and inter-site interactions
between the electrons on  the dots are strong and their coupling  to their
environmental  electron baths relatively weak. Such systems can be used
to probe the effects of strong local electron correlation, such as the Kondo effect,
in great detail.\cite{GSMAMK98,COK98,WFFETK00,KAGGKS04,AGKK05} 
Experimentally it is possible to vary many of the 
parameters in these nanoscale systems in a controlled way; for example, the energy levels
on the dots can be changed through the application of individual gate
voltages to the 
dots making it possible to investigate different parameter regimes.
 Strong correlation behavior in steady
state non-equilibrium conditions can be examined by applying bias voltages to the individual
dots and then  measuring the electron transport through the dots. Borda et
al.\cite{BZHHD03}  have drawn attention to the situation of a singly occupied
double quantum dot where the spin and inter-dot fluctuations are
degenerate. The inter-dot charge fluctuations can be interpreted 
as pseudospin fluctuations,  the occupation of one dot by a single
electron corresponding to an `up' pseudospin, and the single occupation 
of the other dot to a `down'  pseudospin.
On the basis of scaling equations for an effective Kondo model
it was concluded that, in this regime, a new symmetry would emerge on a low energy scale  
between the spin and pseudospin fluctuations, such that the low energy behavior
could be described by an effective model with SU(4) symmetry.\cite{BZHHD03,LS03,LSB04,Zar06,LSL07}
\par

Recently it has proved possible to realize this situation experimentally using
two capacitively coupled dots,\cite{HHWK08,AKRCKSOG13,KAWMRKSZG13pre} and to measure the response
to an effective pseudospin field by changing the levels on the dots. The
conductance of the  electrons through the individual dots has also been
measured, offering the potential to examine the theoretical predictions in
detail. One  technique for calculating the low energy behavior
is via the determination of the renormalized parameters which specify the
effective Hamiltonian in this regime. These can be determined from an analysis
of the low energy fixed point in a numerical renormalization group\cite{Wil75,BCP08} (NRG)
calculation.\cite{HOM04,NCH12b} Once these have been determined several
response functions, such as the spin and charge susceptibilities at zero
temperature and the linear conductance through the dots, can be calculated
from exact expressions for these quantities in terms of these renormalized
parameters. By comparing with exact Bethe ansatz results it has been  shown that very accurate
numerical results can be obtained from these calculations.\cite{HOM04,Hew05,HBK06}
Furthermore leading order corrections to some of these results can
be determined exactly using these parameters within a renormalized
 perturbation theory (RPT).\cite{Hew93} We use this technique in this paper
to examine the circumstances in which the low energy behavior could correspond
to an SU(4) model due to degenerate  spin and inter-dot (orbital)
fluctuations. We calculate the spin and orbital susceptibilities and look at
the effect of introducing a magnetic field to suppress the spin fluctuations
and induce a crossover to an  SU(2) pseudospin  Kondo effect. Finally we estimate 
temperature dependence of the linear conductance in terms of temperature
dependent parameters for the quasiparticles, and show that this  approach gives
results in line with recent experimental observations.
 
\section{Model Hamiltonian}
The  Hamiltonian for the double quantum dot can be expressed in the form,  
\begin{equation}
  H=\sum_{i=1,2}(H_{{\rm d},i}+H_{{\rm bath},i}+ H_{{\rm
    c},i}) + H_{12},
\label{h1}
\end{equation}
where $H_{{\rm d},i}$ describes the individual dots, $i=1,2$, $H_{{\rm
    bath},i}$ the baths to which the dots are individually coupled by
a coupling term $H_{{\rm   c},i}$, and $H_{12}$  is the 
interaction between the dots. 
A reasonable approximation is to take the  baths, two for each dot, to be
described by a free electron model,
\begin{equation}
H_{{\rm bath},i}=\sum_{\vk,\alpha,{\sigma}}\epsilon_{\vk} \elcre{\vk}{i,\alpha,\sigma}
\elann{\vk,i,\alpha}{\sigma} 
\end{equation}
where $\alpha=s,d$ (source, drain) and $\epsilon_{\vk}$ is an energy level in a bath,
taken to be independent of $\alpha$, $i$ and $\sigma$.\par

The Hamiltonian describing the dots $H_{{\rm d},i}$ is taken in the form,
\begin{equation}
H_{{\rm d},i}= \sum_{{\sigma}}\epsilon_{d,i,\sigma}\elcre{d}{i,\sigma}
 \elann{d}{i,\sigma} 
 +  U_i n_{d,i,\uparrow}n_{d,i,\downarrow},
\end{equation}
where $\epsilon_{d,i,\sigma}$ is the level position on dot $i$ in a magnetic
field $h$, $\epsilon_{d,i,\sigma}=\epsilon_{d,i}-\sigma h$,  relative to
the chemical potential $\mu_i$, and $U_i$ is the intra-dot interaction.
It will be useful to introduce an analogous pseudospin field $h_{\rm ps}$
by writing $\epsilon_{d,1}=\bar\epsilon_d-h_{\rm ps}$
and  $\epsilon_{d,2}=\bar\epsilon_d+h_{\rm ps}$,  where
$\bar\epsilon_d=0.5(\epsilon_{d,1}+\epsilon_{d,2})$.\par

The coupling of the dots to the leads is described by a hybridization term,
\begin{eqnarray}
H_{{\rm c},i}
=\sum_{\vk,\alpha,{\sigma}}V_{\vk,i,\alpha}(\elcre{\vk}{i,\alpha,\sigma}\elann{d}{i,\sigma} + \hc).
\end{eqnarray}
We will assume no energy dependence of the matrix elements but allow them to
differ in the different channels. We define the widths $\Gamma_{i,\alpha}=\pi
V_{i,\alpha}^2\rho_c(0)$ with the conduction electrons density of states
$\rho_c$ as the constant energy scale for hybridization, and
their sum, $\Gamma_{i}=\sum_{\alpha} \Gamma_{i,\alpha}$. For transport close
to equilibrium only the  combination
$V_{i,s}\elcre{\vk}{i,s,\sigma}+V_{i,d}\elcre{\vk}{i,d,\sigma}$ 
couples to the dot states. We can therefore simplify the problem to two dots
and two itinerant channels.\par

Finally for the coupling of the dots we assume a hopping term $t$
and a repulsive interaction between the charges on each dot $U_{12}$,
\begin{equation}
  H_{12}=t\sum_\sigma(\elcre{d}{1,\sigma}\elann{d}{2,\sigma} + \hc)+
 U_{12}\sum_{\sigma,\sigma'} n_{d,1,\sigma}n_{d,2,\sigma'}.
\end{equation}
To get an idea of the order of magnitude of these parameters we
quote values estimated in recent experimental work:\cite{AKRCKSOG13}
 $U_1\approx 1.2$meV, $U_2\approx 1.5$meV, $U_{12}\approx 0.1$meV,
 $\Gamma_1,\Gamma_2\approx 0.005-0.02$meV and $t\sim 0$.
Due to the very small value of the hopping term $t$ we will neglect this
term in the calculations presented here.\par
\begin{figure}[!htbp]
\includegraphics[width=0.80\linewidth]{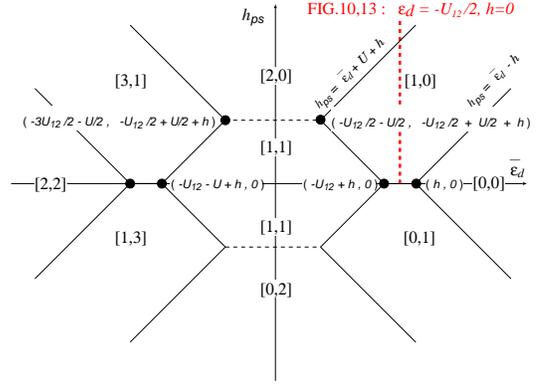}
\caption
{(Color online) The ground state electron configurations of the isolated double dot
in $(\bar\epsilon_d,h_{\rm ps})$ plane at a magnetic  field $h$ $(\ge 0)$. The
meaning of $[n_{d,1},n_{d,2}]$ in the figure is the number of electrons on  dot
$i$ $(=1,2)$ is $n_{d,i}$.}
\label{diag1}
\end{figure}

The ground state electron configurations for the isolated double dot
system for the Hamiltonian Eq.\eqref{h1} with the
occupation numbers $[n_{d,1},n_{d,2}]$ as functions of onsite energy $\bar\epsilon_d$ and pseudospin field
$h_{\rm ps}$ are shown in Fig.~\ref{diag1}.\cite{WFEFTK02}

\noindent
Note that in an experiment $\epsilon_d$ can be tuned via a gate
voltages.\cite{HHWK08,AKRCKSOG13,KAWMRKSZG13pre} 

We now consider an effective model in the low energy regime in terms of
renormalized parameters.\par

\section{The low energy effective model}
The low energy fixed point of this model corresponds to a  Fermi liquid
theory and we can, therefore, assume that the self-energy
$\Sigma_{i,\sigma}(\omega)$ for the single electron Green's function
on dot $i$ is non-singular at $\omega=0$. We can hence describe the low
energy behavior in terms of well defined quasiparticles and their
interactions. We have shown in earlier work \cite{HOM04,NCH12b} in the absence
of a magnetic field that these quasiparticles
can be taken to correspond to renormalized versions of the parameters
that specify the bare  model, $\tilde\varepsilon_{d,i,\sigma}$,
$\tilde\Gamma_{i}$, $\tilde U_i$ and $\tilde
U_{12}$. The Hamiltonian describing the low energy fixed point and the leading
irrelevant correction terms then has the same form as the original model,
but with the bare parameters replaced by the renormalized ones. 
The interaction terms have to be normal ordered as they only come into
play when two or more single particle excitations are created from the
interacting ground state. In the presence of a magnetic field the inter-dot
interaction in the low energy effective Hamiltonian has to be 
generalized to the form,
\begin{equation} \sum_{\sigma,\sigma'}\,\tilde
U_{12}^{\sigma,\sigma'}\! :n_{d,1,\sigma} n_{d,2,\sigma'}:,
\end{equation} 
to allow for the fact that the quasiparticle interactions can be
spin-dependent, where the brackets $:\hat O :$ indicate a normal ordering of
the operator $\hat O $.\par

More generally a renormalized form of perturbation theory can be formulated in
terms of these quasiparticles \cite{Hew93,Hew01,Hew06} in which all
interaction terms of the bare model are included. This requires the explicit
taking into account of counter terms to avoid overcounting renormalization
effects which have already been included in the use of renormalized
parameters. For simplicity we will assume the dots to be identical, apart from the energy
levels $\varepsilon_{i,\sigma}$,  so $\Gamma_{i}=\Gamma$ and 
$U_{i}=U$ and the corresponding renormalized parameters will be taken to be
independent of $i$.\par

Before calculating the renormalized parameters we consider a number of
quantities that can be expressed exactly in terms of these parameters. The linear 
coefficient of the specific heat coefficient $\gamma$ due to the dots is
independent of the quasiparticle interactions, as expected in a Fermi 
liquid theory, and is given by
\begin{equation}
\gamma={\pi^2\sum_{i,\sigma}\tilde\rho_{i,\sigma}(0)\over 3},
\end{equation}
where $\tilde\rho_{i,\sigma}(\omega)$ is the free 
quasiparticle density of states per single spin and channel,
\begin{equation}
\tilde \rho_{i,\sigma}(\omega)={\tilde\Gamma/\pi\over
  (\omega-\tilde\varepsilon_{d,i,\sigma})^2 +\tilde\Gamma^2}.
\label{fqpdos}
\end{equation}
Here $\tilde\Gamma=z\Gamma$, where $z$ is quasiparticle weight factor.
The phase shift $\delta_{i,\sigma}$ per spin on the dot connected to channel $i$
is given by the Friedel sum rule,
\begin{equation}
 \delta_{i,\sigma}=\frac{\pi}{2}-{\rm tan}^{-1}\left
({\varepsilon_{d,i,\sigma}+\Sigma_{i,\sigma}(0)\over \Gamma}\right),
\label{fsr}
\end{equation}
and equivalently in terms of the renormalized parameters,
\begin{equation}
 \delta_{i,\sigma}=\frac{\pi}{2}-{\rm tan}^{-1}\left
({\tilde\varepsilon_{d,i,\sigma}\over \tilde\Gamma}\right).
\label{qpocc}
\end{equation}
The total occupation of the impurity sites $n_{d,\rm tot}$ is given by
$n_{d,\rm tot}=\sum_{i,\sigma}
n_{d,i,\sigma}=\sum_{i,\sigma}\delta_{i,\sigma}/\pi$ at  
$T=0$. These expressions in terms of renormalized parameters already allow us
to draw first conclusions about the occurrence of an emergent SU(4)
symmetry. If, in the absence of an applied magnetic field, 
$\tilde\varepsilon_{d,i}= \tilde\varepsilon_{d}=\tilde\Gamma$ for $i=1,2$ we have
level degeneracy on the dots $\tilde\varepsilon_{d,1}= 
\tilde\varepsilon_{d,2}$, so
$\tilde\rho_{1}(0)=\tilde\rho_{2}(0)=\tilde\rho(0)=1/2\pi\tilde\Gamma$,  and the phase
shifts per spin per dot are all  equal to  $\pi/4$.
Hence at the free quasiparticle level the system in this regime has SU(4)
symmetry. However, the quasiparticle interaction terms play an important role
in determining the low energy behavior. They correspond to the leading
correction terms to the fixed point, and so the low energy model only has
SU(4) symmetry if this symmetry is retained when these terms are included.\par

Other exact equations are for the total charge susceptibility,\cite{NCH12b} 
\begin{equation}
\chi_c=\sum_{\sigma}[\tilde\eta_{c,1,\sigma} \tilde\rho_{1,\sigma}(0)
+\tilde\eta_{c,2,\sigma} \tilde\rho_{2,\sigma}(0)],
\label{chic}
\end{equation}
where the term $\tilde\eta_{c,i,\sigma}$ takes into account the quasiparticle
interactions and is given by
\begin{equation}
\tilde\eta_{c,i,\sigma}=1-\tilde U\tilde\rho_{i,-\sigma}(0) -\sum_{i'\ne i,\sigma'}\tilde
U_{12}^{\sigma,\sigma'}\tilde\rho_{i',\sigma'}(0).
\end{equation}
\par
In the case with level degeneracy on the dots, and no external magnetic field,
we can expect the total charge susceptibility to be negligible if
$U/\pi\Gamma\gg 1$ and  $U_{12}/\pi\Gamma\gg 1$. This is because double
occupancy on a single dot is inhibited by the large value of $U$
and double occupation of the two dots with one electron on each dot  is
inhibited by the large value of $U_{12}$. 
Equating the total charge susceptibility to zero  at this degeneracy point
gives a relation between the renormalized parameters,
\begin{equation}
\tilde U +2\tilde
U_{12}=2\pi\tilde\Gamma.
\label{rel1}
\end{equation}
Away from this degeneracy point, if the ground state of the system has on
average one electron on the two dots, and $U/\pi\Gamma\gg 1$ and
$U_{12}/\pi\Gamma\gg 1$, then we still expect the charge susceptibility to be
negligible and we get a more general condition,
\begin{equation}
\sum_{i=1,2} \tilde\rho_i(0)[1-
\tilde U\tilde\rho_i(0)]
-4\tilde U_{12}\tilde\rho_1(0)\tilde\rho_2(0)
=0
\label{rel2}
\end{equation}
The total spin susceptibility, $\chi_{\rm s}=\sum_i d m_i/dh$,
$m_i=(n_{d,i,\uparrow}-n_{d,i,\downarrow})/2$,  of the two dots is given
by\cite{NCH12b} 
\begin{equation}
\chi_{\rm s}=\sum_{\sigma}[\tilde\eta_{s,1,\sigma}
\tilde\rho_{1,\sigma}(0)+\tilde\eta_{s,2,\sigma} \tilde\rho_{2,\sigma}(0)], 
\label{chis}
\end{equation}
where
\begin{equation}
\tilde\eta_{s,i,\sigma}=1+\tilde U\tilde\rho_{i,-\sigma}(0),
\end{equation}
and the pseudospin susceptibility, $\chi_{\rm ps}=d\, m_{\rm ps}/d h_{\rm ps}$,
$m_{\rm ps}=(n_{d,1}-n_{d,2})/2$,   by
\begin{equation}
\chi_{\rm ps}=\sum_{\sigma}[\tilde\eta_{ps,1,\sigma}
\tilde\rho_{1,\sigma}(0)+\tilde\eta_{ps,2,\sigma} \tilde\rho_{2,\sigma}(0)], 
\label{chisp}
\end{equation}
where $\tilde\eta_{ps,i,\sigma}$  is given by
\begin{equation}
\tilde\eta_{ps,i,\sigma}=1-\tilde U\tilde\rho_{i,-\sigma}(0) +\sum_{i'\ne i,\sigma'}\tilde
U_{12}^{\sigma,\sigma'}\tilde\rho_{i',\sigma'}(0).
\end{equation}
At the degeneracy point in the absence of a magnetic field these become 
\begin{equation}
\chi_{\rm s}={1\over \pi\tilde\Gamma}
\left (1+{\tilde U\over 2\pi\tilde\Gamma}\right),
\label{dchis}
\end{equation}
and for the pseudospin
\begin{equation}
\chi_{\rm ps}={1\over \pi\tilde\Gamma}
\left (1+{2\tilde U_{12}-\tilde U\over 2\pi\tilde\Gamma}\right).
\label{dchips}
\end{equation}
For SU(4) symmetry of the effective  Hamiltonian with renormalized
parameters determining the low energy behavior at this degeneracy point we
require $\tilde U_{12}=\tilde U$, which as expected makes the spin and
pseudospin susceptibilities equal. 
From Eq.~(\ref{rel1}) this implies $\tilde U_{12}=2\pi\tilde\Gamma/3=\tilde
U$, giving the known  Wilson ratio, $W_{\rm s}=\pi^2\chi_{\rm s}/(3\gamma)$, for an SU(4)
Kondo model of 4/3. We have only one energy scale in this case 
which will be the Kondo temperature  $T_{\rm K}$ for the SU(4) Kondo model
which we can define by the relation 
\begin{equation}
4T_{\rm K}={1\over \tilde\rho(0)}=\pi\tilde\Gamma
\Big(1+\frac{\tilde\epsilon_{d}^2}{\tilde\Gamma^2}\Big)=2\pi\tilde\Gamma,  
\label{eq:TKscale}
\end{equation}
where for the last equation the degeneracy point was assumed.

If we raise the spin degeneracy by an applied  magnetic field
but keep the average electron occupation on each dot as $1/2$,
and $\rho_{1,\sigma}(0)=\rho_{2,\sigma}(0)$, then eventually in a large magnetic
field we will be left with only one spin type on each dot. We take this to
correspond to spin up so that in this limit, $\rho_{i,\downarrow}(0)\to 0$. For
magnetic field energies small compared with both $U_{12}$ and
$U$,  we can still equate the charge susceptibility to zero,
which would imply  $\rho_{i,\uparrow}(0)\to 1/\pi\tilde\Gamma$. 
 There is then no enhancement of the spin susceptibility, but an enhancement
of the   pseudospin susceptibility  by a factor $1+\tilde
  U_{12}/\pi\tilde\Gamma$ corresponding to a pseudospin Kondo effect.
For $U_{12}\gg \pi\Gamma$, we have $\tilde   U_{12}/\pi\tilde\Gamma\to 1$
giving the SU(2) pseudospin Wilson ratio
$W_{\rm ps}=\pi^2\chi_{\rm ps}/(3\gamma)=2$.

To test these relations, and more generally evaluate spin
and pseudospin susceptibilities as a function of the energy levels on the dot
and applied magnetic field, we need to calculate the
renormalized parameters. We have described in earlier work how these can be
deduced from an analysis of the low energy fixed point in a numerical
renormalization group (NRG) calculation.\cite{HOM04,NCH12b} We apply the
method to the model being investigated here and describe the results of these
calculations in detail in the next section. We use $\Lambda=6$ (this
comparatively
large value gives accurate estimates for the renormalized parameters as can
be checked in the case of a single impurity model) and
typically retain $4000$ states in our NRG calculations. 

\section{Calculation of the renormalized parameters}
\begin{figure}[!htbp]
\includegraphics[width=0.80\linewidth]{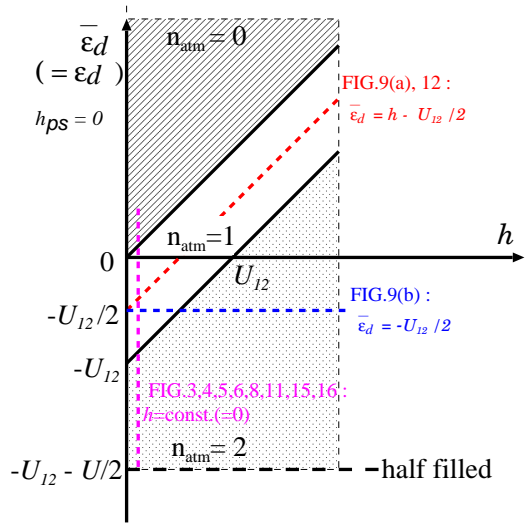}
\caption
{(Color online) The ground state electron occupation numbers in the isolated double dot in
  $(h,\bar\epsilon_d)$ plane at zero pseudospin field $h_{\rm ps}=0$.}
\label{diag2}
\end{figure}

We first of all test the hypothesis that the two dot model with
degenerate levels and a total occupation of the two dots $n_{d, \rm tot}\sim 1$
has an emergent SU(4) low energy fixed point in a regime where fluctuations
in the total charge on the two dots are suppressed, 
$U/\pi\Gamma\gg 1$ and $U_{12}/\pi\Gamma\gg1$. We first analyze the situation
where the onsite energy $\epsilon_d$ is varied. This corresponds to the line
along $h=0$ in Fig.~\ref{diag2}, which can serve as a guideline.

\begin{figure}[!htbp]
\includegraphics[width=0.80\linewidth]{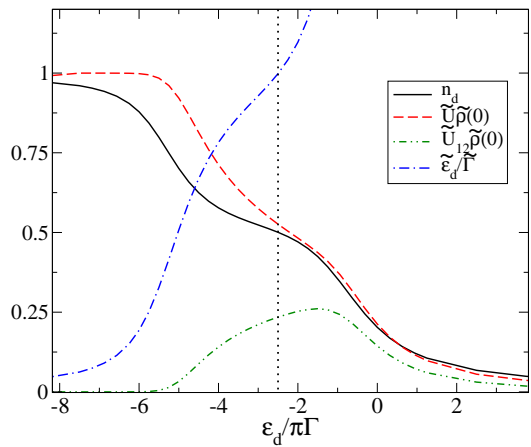}
\caption
{(Color online) A plot of the  occupation number on a single dot $n_{d}$, $\tilde
  U\tilde\rho(0)$, $\tilde U_{12}\tilde\rho(0)$ and
$\tilde\varepsilon_d/\tilde\Gamma$  as a function of $\varepsilon_d/\pi\Gamma$
for the model with $U/\pi\Gamma=20$, $U_{12}/\pi\Gamma=5$ and
  $\pi\Gamma=0.01$.
The vertical dotted line corresponds to $\varepsilon_d=-U_{12}/2$.  }
\label{rp1}
\end{figure}

\noindent
We consider the cases with parameters similar to those quoted in
experiment.\cite{AKRCKSOG13} Many of the following results are for
$U/\pi\Gamma=20$, $U_{12}/\pi\Gamma=5$, and we take $\pi\Gamma=0.01$ in all of the
calculations presented here. We consider the case first of all with
$\epsilon_{d,1}=\epsilon_{d,2}=\epsilon_{d}$ ($h_{\rm ps}=0$). In  Fig.~\ref{rp1} we plot
$n_{d,i}=n_{d}$, the occupation number on each dot, and 
the combinations of renormalized parameters $\tilde U\tilde\rho(0)$, $\tilde
U_{12}\tilde\rho(0)$ and $\tilde\varepsilon_d/\tilde\Gamma$  as a function of
the level position on the dots $\varepsilon_d/\pi\Gamma$.

\noindent
Over this range the total occupation of the 
two dots varies from a regime with $n_{d,\rm tot}\sim 2$, 
where $\tilde\varepsilon_d/\tilde\Gamma\to 0$, $\tilde U\tilde\rho(0)\to 1$,
corresponding to a spin Kondo regime on each dot, to a low density
weakly correlated regime $n_{d,\rm tot}\sim 0$, where both $\tilde U\tilde\rho(0)\to 0$
and $\tilde U_{12}\tilde\rho(0)\to 0$. For $\varepsilon_d/\pi\Gamma\sim -2.5$,
which corresponds to $\varepsilon_d\sim -U_{12}/2$, there is a region where $n_{d,\rm tot}\sim 1$
and approximate degeneracy of the spin and inter-dot excitations.

\begin{figure}[!htbp]
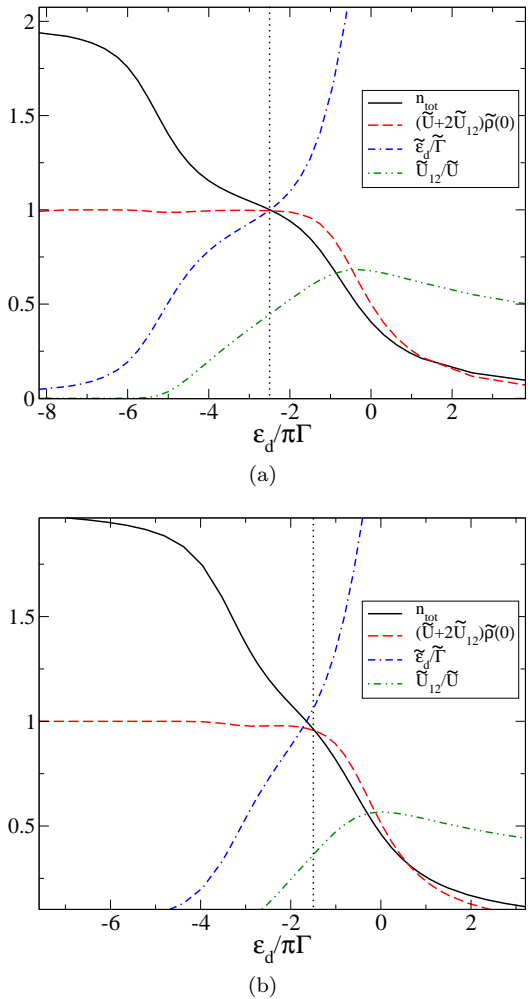
 
\subfigure[]{\includegraphics[width=0.80\linewidth]{figure4.eps}}
\vspace{0.2cm}
\subfigure[]{\includegraphics[width=0.80\linewidth]{figure6.eps}}
\caption
{(Color online) A plot the total occupation number on the dots $n_{d,\rm tot}$,
  $\tilde\rho(0)(\tilde U+2\tilde U_{12})$, $\tilde\varepsilon_d/\tilde\Gamma$
  and $\tilde U_{12}/\tilde U$  as a function of $\varepsilon_d/\pi\Gamma$ 
for (a) the parameter set given in Fig.~\ref{rp1} $U/\pi\Gamma=20$,
$U_{12}/\pi\Gamma=5$ and (b) $U/\pi\Gamma=12$,
$U_{12}/\pi\Gamma=3$, where $\pi\Gamma=0.01$.
The vertical dotted lines correspond to $\varepsilon_d=-U_{12}/2$.}
\label{test}
\end{figure}

To check some of the predicted relations between the renormalized parameters
we plot the combination $\tilde\rho(0)(\tilde U+2\tilde U_{12})$, relevant for
the charge susceptibility in Eq. (\ref{chic}), and the ratios
$\tilde\varepsilon_d/\tilde\Gamma$, and $\tilde U_{12}/\tilde 
U$ in Fig.~\ref{test}~(a). The total occupation of the two dots  $n_{d,\rm tot}$
is also shown.

\noindent
For  $\varepsilon_d/\pi\Gamma< -2.0$ it can be seen
that the combination  $\tilde\rho(0)(\tilde U+2\tilde U_{12})$ is
very close to the value of 1, which from  Eq.~(\ref{rel2}) implies  a localized regime where the
total charge susceptibility of the two dots is negligible,
and the fluctuations in the total charge have been almost completely
suppressed. This regime includes the point of complete degeneracy between the
spin and inter-dot charge fluctuations, where $\tilde\varepsilon_d/\tilde\Gamma=1$ and  $n_{d,\rm
  tot}=1$ so that, to a good approximation, all three curves have a common point
of intersection, as can be seen clearly in Fig.~\ref{test}~(a). 
This point to a good approximation corresponds to  $\epsilon_d=-U_{12}/2$.

If this degeneracy point corresponded to an SU(4) symmetry for the low
energy excitations then we would expect  the ratio $\tilde U_{12}/\tilde
U$ to pass through this same point giving  $\tilde U_{12}=\tilde
U$. However, it is of the order 0.45, substantially smaller than
1. The ratio is closer to 1 than that of the bare values $U_{12}/U=0.25$, and
hence there is a flow towards the symmetry point, which is however not reached
for experimentally relevant parameters. The values, $\tilde\rho(0)\tilde  
U_{12}=0.23$ and  $\tilde\rho(0)\tilde U=0.54$, at the degeneracy point 
give a Wilson ratio $W_{\rm s}$  for the spin, $W_{\rm s}=1+\tilde
U\tilde\rho(0)=1.54$, and for the pseudospins $W_{\rm ps} =1+2\tilde
U_{12}\tilde\rho(0)-\tilde U\tilde\rho(0)=0.93$.  As these differ we do not
have  SU(4) symmetry at the degeneracy point for this parameter set,  SU(4)
symmetry would require  $W_{\rm s}=W_{\rm ps}=4/3$.

\begin{figure}[!htbp] 
\includegraphics[width=0.80\linewidth]{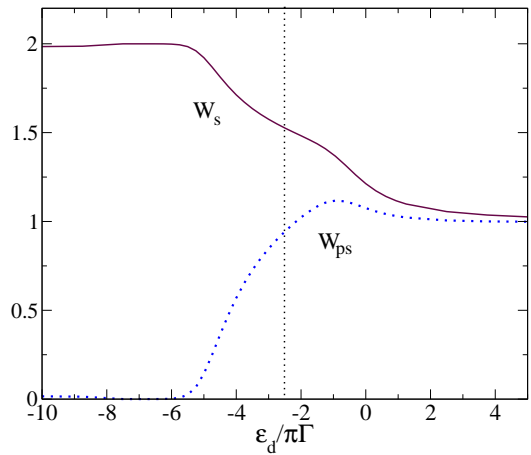}
\caption
{(Color online) A plot of the Wilson ratios for the spin excitations $W_{\rm s}$ and pseudospin
  excitations $W_{\rm ps}$ for the double dot 
as a function of $\epsilon_d/\pi\Gamma$
for the parameter set given in Fig.~\ref{rp1}.
The vertical dotted line corresponds to $\varepsilon_d=-U_{12}/2$.
}
\label{wr_vs_ed}
\end{figure}

How the two Wilson ratios, $W_{\rm s}$  and $W_{\rm ps}$, vary with $\epsilon_d$
for the same parameter set as in Fig.~\ref{rp1}, is shown in
Fig.~\ref{wr_vs_ed}. 

\noindent
When $-U/2<\epsilon_d<-U_{12}$, the interdot interaction 
$U_{12}$ plays no significant role and $\tilde U_{12}\to 0$. In this regime  $\tilde
U\tilde\rho(0)\to 1$ which  has the effect of suppressing the
pseudospin fluctuations so $W_{\rm ps}\to 0$ [see Eq.~\eqref{dchips}] and at the same
time enhancing the spin Wilson ratio $W_{\rm s}\to 2$. This corresponds to the spin
Kondo limit with a single electron on each dot. As $\epsilon_d$ is increased
from  $\epsilon_d=-U_{12}$  the value of $\tilde U_{12}$ increases and $\tilde
U$ decreases, which has the effect of enhancing $W_{\rm ps}$ and 
reducing $W_{\rm s}$, but as long as the bare interactions obey $U<U_{12}$,
$W_{\rm ps}<W_{\rm s}$. As the level $\epsilon_d$ on the dots  passes above the Fermi
level the interaction terms play very little role and both $W_{\rm ps}$ and
$W_{\rm s}$ asymptotically approach the value 1 corresponding to non-interacting
quasiparticles. 

We take a closer look at some of the results for different parameters and
compare them with the study for the case $U/\pi\Gamma=20$, 
$U_{12}/\pi\Gamma=5$. For  $U/\pi\Gamma=12$, $U_{12}/\pi\Gamma=6$ the 
results were very similar to those presented in Fig.~\ref{test}~(a) but with a slightly increased value of
$\tilde\rho(0)\tilde U_{12}$ due to the relatively larger value of $U_{12}$
compared with $U$. However, for the parameter set  $U/\pi\Gamma=12$,
$U_{12}/\pi\Gamma=3$, with a relatively smaller value of  $U_{12}$ there are
some qualitative differences.  This is shown in
Fig.~\ref{test}~(b), which can be compared directly with the corresponding plot in
Fig.~\ref{test}~(a). It can be seen  that, due to the smaller value of  $U_{12}/\Gamma$,
the  fluctuations of the total charge on the two dots  are not suppressed completely so that
the value of $\tilde\rho(0)(\tilde U+2\tilde U_{12})$ is slightly less than 1
at the spin-pseudospin degeneracy point so that it falls below this point.\par

\begin{figure}[!htbp]
\includegraphics[width=0.88\linewidth]{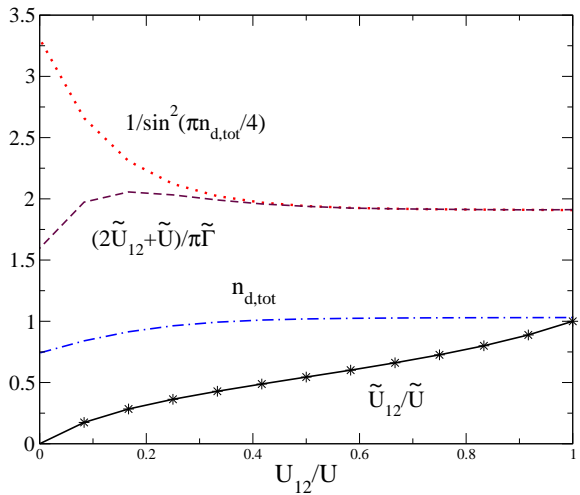}
\caption
{(Color online) A plot of $n_{d,\rm tot}$, $(2\tilde U_{12}+\tilde
  U)/\pi\tilde\Gamma$, $1/{\rm sin}^2(\pi n_{d,\rm tot}/4)$ and $\tilde
  U_{12}/\tilde U$ as a function of $U_{12}/U$ for $\epsilon_d=-U_{12}/2$.
}
\label{extratest}
\end{figure}

To test more generally the possibility of an emergent SU(4) low energy 
fixed point in the regime $U/\pi\Gamma\gg 1$,  $U_{12}/\pi\Gamma\gg 1$
with $n_{d,\rm tot}\sim 1$, we have calculated the renormalized parameters
$ \tilde U$, $\tilde U_{12}$, $\tilde\Gamma$ and $\tilde \epsilon_d$ as a function of
$U_{12}$ for the
case $U/\pi\Gamma=12$ with $\epsilon_d=-U_{12}/2$.  The results for $n_{d,\rm tot}$, $(2\tilde U_{12}+\tilde
  U)/\pi\tilde\Gamma$, $1/{\rm sin}^2(\pi n_{d,\rm tot}/4)$ and $\tilde
  U_{12}/\tilde U$ are shown as a function of $U_{12}/U$ in
  Fig.~\ref{extratest}.  The condition $(2\tilde U_{12}+\tilde
  U)/\pi\tilde\Gamma=1/{\rm sin}^2(\pi n_{d,\rm tot}/4)$, which holds to a good
  approximation for $U_{12}/U>0.4$ implies that the total charge
  susceptibility  is negligible and  $n_{d,\rm tot}\sim 1$.
We find the condition $\tilde U_{12}/\tilde U=1$ for a low energy SU(4) fixed
point  is only satisfied when $U_{12}=U$, ie. only if we have SU(4) symmetry already
for the `bare' model. At the SU(4) fixed point with the condition that the
charge susceptibility is set to zero we predict
\begin{equation}
{\tilde U\over\pi\tilde\Gamma}={\tilde U_{12}\over\pi\tilde\Gamma}=
{1\over 3{\rm sin}^2(\pi n_{d,\rm tot}/4)},
\label{SU4}
\end{equation}
which is satisfied precisely in the results in Fig. \ref{extratest}. 
With the choice $\epsilon_d=-U_{12}/2$ for  $U_{12}/\pi\Gamma\gg 1$
we have $n_{\rm tot}\sim 1$ but not precisely equal to 1. In an earlier
study of the SU(4) version of this model\cite{NCH10a} we calculated the renormalized
parameters keeping $n_{d,\rm tot}$ strictly equal to 1. In that case we obtained
the result $\tilde U/\pi\tilde\Gamma=0.66665$  asymptotically for large
$U$ in very accurate agreement with the prediction from Eq. (\ref{SU4}) 
$\tilde U/\pi\tilde\Gamma=2/3$.\par
\begin{figure}[!htbp]
\includegraphics[width=0.82\linewidth]{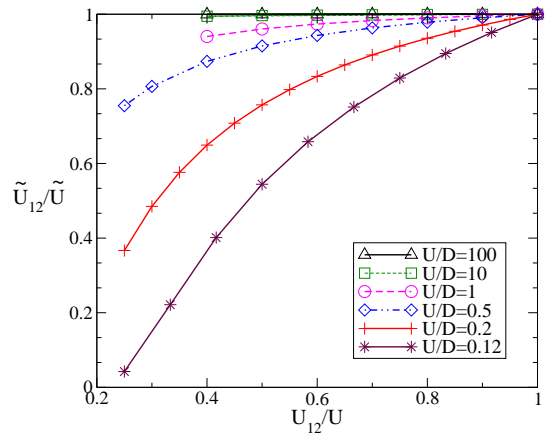}
\caption
{(Color online) A plot of  $\tilde
  U_{12}/\tilde U$ as a function of $U_{12}/U$ for $\epsilon_d=-3\pi\Gamma$ and
$U/D=0.12,0.2,0.5,1,10,100$.
}
\label{extratest2}
\end{figure}

\noindent
We get a very similar picture to that shown in Fig. \ref{extratest}
if we take a larger value of $U$,  $U/\pi\Gamma=20$, $\epsilon_d=-U_{12}/2$
and vary $U_{12}$. We find  $\tilde U_{12}/\tilde U=1$ only when  $U_{12}=U$.
For a given ratio $U_{12}/U$,
the ratio of $\tilde U_{12}/\tilde U$ is observed to be larger in the range $ U_{12}<U$  when we increase the value 
of  $U$ from  $U/\pi\Gamma=12$ to $U/\pi\Gamma=20$. The question arises as to whether
the ratio  $\tilde U_{12}/\tilde U$ would approach the value 1 in this range if we
increase the value of $U$ still further. 
 As $U/\pi\Gamma>20=0.2/D$ ($D=1$, $\pi\Gamma=0.01$), then this would mean taking values of $U$ 
comparable with the half-bandwidth $D$. With $U\sim D$ and $\epsilon_d=-U_{12}/2$
in the range $U_{12}\to U$, the ground state impurity level will become far removed from the Fermi level
resulting in an renormalized energy scale $T^*\to 0$.
 To investigate the larger $U$ regime, 
therefore, we take a fixed value for  $\epsilon_d$, just below the Fermi level
 $\epsilon_d=-3\pi\Gamma$. We know from the Schrieffer-Wolff transformation
that for $U=U_{12}>D$ that the model in this regime is equivalent to an SU(4)
Coqblin-Schrieffer model.\cite{hewson,NCH10a} 
The question arises as to whether this mapping still holds if $U_{12}<U$, and
if it also applies to a parameters with $U<D$. In Fig.~\ref{extratest2} we give the results
for $\tilde   U_{12}/\tilde U$ as a function of $U_{12}/U$  for  $\epsilon_d=-3\pi\Gamma$ and
$U/D=0.12,0.2,0.5,1,10,100$.  We see that for values of $U/D\le 1$, 
we get the SU(4) symmetric case  with  $\tilde
  U_{12}/\tilde U=1$ only if $U_{12}=U$. However, for $U/D>1$ we do have a finite range
where  the ratio $\tilde U_{12}/\tilde U$ is almost 1 and so SU(4) symmetry is
effectively realized. However, this appears
to be strictly only the case only if $U_{12}$ is also greater than the
half-band width $D$.  

\begin{figure}[!htbp]
\includegraphics[width=0.80\linewidth]{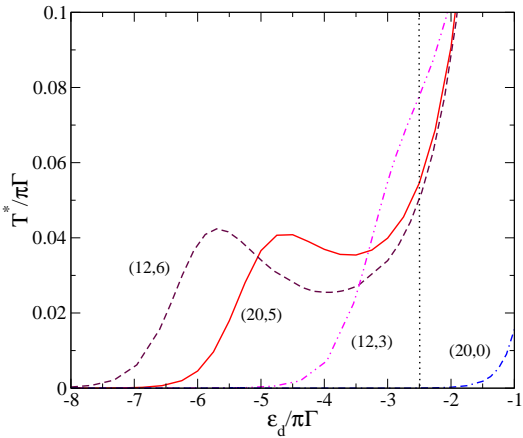}
\caption
{(Color online) A plot of $T^*/\pi\Gamma$   as a function of $\varepsilon_d/\pi\Gamma$
for different parameter sets labeled by  the values   ($U/\pi\Gamma$, $U_{12}/\pi\Gamma$).
The vertical dotted line corresponds to $\varepsilon_d=-U_{12}/2$ for case  (20,5).
}
\label{Tstar}
\end{figure}

\noindent
As in Eq.~(\ref{eq:TKscale}) we can define an energy scale $T^*$ via
$4T^*=1/\tilde\rho(0)$. 
It has the property $T^*\to T_{\rm K}$, where  $T_{\rm K}$ is the spin Kondo
temperature in the range where $n_{d,i}\sim 1$, defined such that 
$\chi_{\rm s}=1/4T_{\rm K}$ for a single dot.  A more significant difference
between the results for the different parameter sets discussed above in
Fig.~\ref{test}(a)  and (b) can be seen in Fig.~\ref{Tstar} where we
plot $T^*/\pi\Gamma$  as a function of $\varepsilon_d/\pi\Gamma$.

\noindent
For the parameter set $(U/\pi\Gamma, U_{12}/\pi\Gamma)=(20,5)$ given in
Fig.~\ref{rp1} we see that $T^*$ 
has a local minimum at  $\varepsilon_d/\pi\Gamma\sim -3.4$ with
$T^*/\pi\Gamma\sim 0.035$. There is an  even more extended and marked minimum
for the parameter set $(12,6)$  shown
as the dashed curve in Fig.~\ref{Tstar}. These two results are in marked
contrast to the result for $T^*/\pi\Gamma$ for the case $(12,3)$ with the smaller value
of $U_{12}/\pi\Gamma$,  [see Fig.~\ref{test}(b)], which has no minimum or even a plateau region.
 
The occurrence of such local minima in $T^{*}$ can be accounted for by
considering the effective model that results from a Schrieffer-Wolff 
transformation on the Hamiltonian Eq.~\eqref{h1} (see the Appendix).  
For $0>\epsilon_{d}>-U_{12}$ and $U\to\infty$, fluctuations between the
fourfold-degenerate atomic groundstates with $n_{d}=1$  
are mediated by both the excited two-particle and unoccupied states, so that
the resulting effective model features a ``pseudospin'' exchange coupling
$J_{\rm ps}\sim -V^{2}U_{12}/[\epsilon_{d}(\epsilon_{d}+U_{12})]$, which is
minimized at the degeneracy point when $\epsilon_{d}=-U_{12}/2$.  This local
minimum in $J_{\rm ps}$ is sufficient to explain the local minimum in the
nominal pseudospin ``Kondo temperature'', $T_{\rm ps}\sim T^{*}\sim
\e^{-1/2\rho_{c}J_{\rm ps}}$, seen in Fig.~\ref{Tstar} for the parameter sets with
$U_{12}/\pi\Gamma=6$ and $U_{12}/\pi\Gamma=12$.  Although the two pseudospin
projections suggest a correspondence in this regime with an SU(2) pseudospin
Kondo model, the spin degrees of freedom modify both the pseudospin Kondo
temperature $T_{\rm ps}$ from its SU(2) value and the location of the minima in
$T^{*}$. The shift of the minimum to the left can be understood qualitatively
by the fact that for large $U$ the spin Kondo coupling
$J_{\rm s}\sim-V^{2}/\epsilon_{d}$ is decreasing on decreasing
$\epsilon_{d}$. Hence, due to the interplay of spin and pseudospin Kondo
effects the minimum shifts to smaller values of $\epsilon_d$.        
Specifically, the minimum for the  parameter set (20,5)
does not correspond to the spin/pseudospin degeneracy point,
which occurs where $\varepsilon_d/\pi\Gamma\sim -2.5$, giving
a value  $T^*/\pi\Gamma\sim 0.055$.  The value of $T^*$ in this regime is very
much greater  than the values of  $T_{\rm K}$  in the spin Kondo regime, 
$\varepsilon_d/\pi\Gamma < -8.0$, which is also shown in Fig.~\ref{Tstar}
for $(20,0)$. \par
    
We can make a comparison of $T^*$ at the degeneracy point  with $T_{\rm K}$  
for an SU(4) model  with $U=U_{12}$ and  $U_{12}/\pi\Gamma=5.0$.\cite{NCH10b}
There are two such SU(4) Kondo models corresponding to the  total occupation numbers,
$n_{d,\rm tot}=1$ and $n_{d,\rm tot}=2$.
For the SU(4)  model with  $n_{d,\rm tot}=1$ we find
$T^*/\pi\Gamma=T_{\rm K}/\pi\Gamma=0.096$, which is
greater than but of the same order 
of magnitude as the value $T^*/\pi\Gamma\sim 0.055$ deduced from
the results in Fig.~\ref{Tstar} at the degeneracy point. The  particle-hole symmetric SU(4) model with $n_{d,\rm
  tot}=2$ has a somewhat lower value of $T^*/\pi\Gamma=T_{\rm
  K}/\pi\Gamma=0.031$. \par

We can estimate the degree of quasiparticle renormalization at the spin/pseudospin
degeneracy point for the parameter set,  $U/\pi\Gamma=20$,
$U_{12}/\pi\Gamma=5$,  by comparing the value  $T^*$ with that for the
corresponding point for the non-interacting system where  $n_{d,\rm tot}=1$. 
At this point, $\epsilon_d/\Gamma=1$ ($n_{d,\rm tot}=1$) which gives $T^*/\pi\Gamma\sim 0.5$.
The degree of renormalization due to the interactions can be estimated from
their ratio $0.5/0.055$, which gives a  renormalization factor  of the order
of 9 in this case.

\section{Results in a field}
\subsection{Crossover as a function of magnetic field $h$ }
\begin{figure}[!htbp]
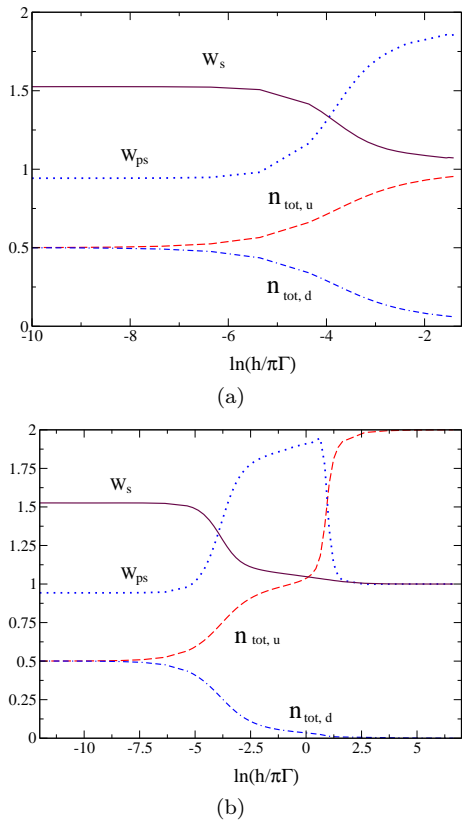

\subfigure[]{\includegraphics[width=0.70\linewidth]{figure8.eps}}
\subfigure[]{\includegraphics[width=0.70\linewidth]{figure9.eps}}
\caption
{(Color online) A plot of the Wilson ratios for the spin and pseudospin degrees of freedom,
  $W_{\rm s}$ and  $W_{\rm ps}$, and the total spin up and spin down occupation
  numbers, $n_{\rm tot,u}$ and $n_{\rm tot,d}$, for the double dot in a
  magnetic field $h$, (a)  with a constraint such that $n_{\rm
    tot}=1$ and (b) without constraint, as a function of
  ${\rm ln}(h/\pi\Gamma)$ for $U/\pi\Gamma=20$, $U_{12}/\pi\Gamma=5$, $\pi\Gamma=0.01$.}
\label{wilson_ratio_h2}
\end{figure}

At the degeneracy point where  the occupation number on each dot $n_{d,i}=0.5$ and
$U\gg \pi\Gamma$ and $U_{12}\gg \pi\Gamma$, we have both spin and
pseudospin fluctuations. Applying a magnetic field at this point will suppress 
the spin fluctuations. With a large enough magnetic field it should be
possible to suppress the spin fluctuations completely  such that there is a
crossover to an SU(2) Kondo fixed point due to the pseudospin fluctuations. If this
proves to be possible experimentally then one could examine the transport
of the two types of pseudospins independently as each is associated with
a single dot only.
The question naturally arises therefore as to how large does the magnetic
field have to be to see this crossover.  To answer this question we have
calculated the renormalized parameters in a magnetic field\cite{HBK06,BH07a} and used them to
deduce the Wilson ratios for the  spin and  pseudospin, $W_{\rm s}$ and
$W_{\rm ps}$. One way of applying the magnetic field is to adjust the mean level
on the dots $\bar\epsilon_d$ such that  $\bar\epsilon_d=h-U_{12}/2$, which,
starting at $\epsilon_d=-U_{12}/2$, will be such as to maintain 
the total occupation of the two dots $n_{\rm d, tot}=1$. This corresponds to
line (1) in Fig.~\ref{diag2}.  The results for this case are shown in
Fig.~\ref{wilson_ratio_h2}~(a) plotted as a function of  ${\rm ln}(h/\pi\Gamma)$
for   $U/\pi\Gamma=20$, $U_{12}/\pi\Gamma=5$, $\pi\Gamma=0.01$. 
We have defined $n_{\rm tot,u}=\sum_i n_{d,i,\uparrow}$ and $n_{\rm
tot,d}=\sum_i n_{d,i,\downarrow}$. 

\noindent
For this parameter set we have $T^*/\pi\Gamma\sim 0.055$ at the degeneracy point corresponding
to  ${\rm ln}(T^*/\pi\Gamma)\sim -2.9$. From  Fig.~\ref{wilson_ratio_h2}~(a)
we can see that the crossover occurs relatively slowly as the magnetic field
is increased but when  $h=T^*$, the pseudospin ratio has risen to a value
$W_{\rm ps}\sim 1.7$ and the spin ratio fallen to $W_{\rm s}\sim 1.14$. At this point
the crossover is well advanced, and so  $T^*$ at the degeneracy point sets
the scale of the crossover with the magnetic field $h$. However, 
one needs larger fields to suppress the spin
fluctuations fully  such that spin ratio $W_{\rm s}$ falls to the value 1 and the
pseudospin ratio $W_{\rm ps}$ reaches the SU(2) Kondo value of 2. From 
Fig.~\ref{wilson_ratio_h2}~(a), we can extract a rough estimate for the polarizing
field ${\rm ln}(h_{\rm pol}/\pi\Gamma)=-1$, $h_{\rm pol}\approx 1.16
\Gamma$. Assuming $\Gamma = 0.01$meV, $h=g\mu_BH/2$ with $|g|=0.44$ for
GaAs,\cite{KAWMRKSZG13pre} the corresponding magnetic field is $H=0.86$T, well
within experimental reach.

A similar crossover behavior is found if the magnetic field is applied at the
degeneracy point without any other adjustment. This corresponds to line (3)
in Fig.~\ref{diag2}. The results  for this case are given in Fig.~\ref{wilson_ratio_h2}~(b)
and the value of $T^*$ sets the scale of the  crossover in this case as well.
There is no constraint in this case to maintain $n_{\rm d, tot}=1$, so 
there is a second crossover when  ${\rm ln}(h/\pi\Gamma)\sim 1$, 
$h/\pi\Gamma\sim 2.7$, which  occurs  when  $h\sim U_{12}/2$. 
When $h>U_{12}/2 $ the interdot interaction no longer plays 
a significant role in determining the occupation numbers on the two dots and 
the two dots become fully polarized such that $n_{1,\uparrow}=n_{2\uparrow}\sim 1$,
and $n_{1\downarrow}=n_{2\downarrow}\sim 0$. Both the spin and pseudospin Kondo
effects are suppressed and the Wilson ratios for both spin and pseudospin
fall to the value 1. 

\subsection{Crossover as a function of pseudospin field $h_{\rm ps}$}
\begin{figure}[!htbp]
\includegraphics[width=0.70\linewidth]{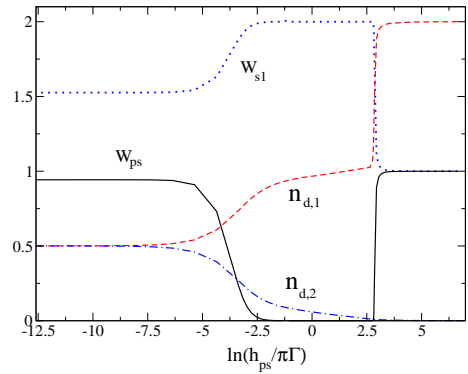}
\caption
{(Color online) A plot of the spin  Wilson ratio $W_{\rm s1}$ on dot 1, the
  pseudospin ratio $W_{\rm ps}$, and the  occupation   numbers for the two dots,
  $n_{\rm d,1}$ and $n_{\rm d,2}$, as a function of ${\rm ln}(h_{\rm ps}/\pi\Gamma)$
for   $U/\pi\Gamma=20$, $U_{12}/\pi\Gamma=5$, $\bar\epsilon_d=-U_{12}/2$, $\pi\Gamma=0.01$ }
\label{wilson_ratio_hps}
\end{figure}

A similar crossover can occur if we change the relative levels on the two 
dots so as to induce an effective field $h_{\rm ps}$ on the pseudospin degrees
of freedom. The results are shown in Fig.~\ref{wilson_ratio_hps} for the
Wilson ratio on dot 1, $W_{\rm s1}$, and the pseudospin ratio $W_{\rm ps}$, together 
with the occupation numbers on the individual dots for the same
parameter set with $\bar\epsilon_d=-U_{12}/2$.

\noindent
The pseudospin field has the effect of suppressing the pseudospin degrees of
freedom leaving the spin degrees of freedom on the dots.
The spin degrees of  freedom, however, depend on the occupation numbers on the
individual dots which also change. When $h_{\rm ps}>T^*$ the pseudospin degrees of
freedom are rapidly suppressed $W_{\rm ps}\to 0$ and the Wilson ratio for the spin
on dot 1 has a plateau region with $W_{\rm s1}\sim 2$. When $h_{\rm ps}$ reaches a
value of the order of $U/2$ (${\rm ln}(U/2\Gamma)={\rm ln}(10)\sim 2.3$) the
occupation number on dot 1 rapidly jumps from the order of 1 to 2. As both
spin states are then occupied on dot 1 the Wilson ratio  $W_{\rm s1}$ fall to the
value 1.

\section{Differential conductance}
A quantity well accessible in experiment is the differential conductance.
The current through dot $i$, $I_i$ is given by the result of Meir and Wingreen,\cite{MW92} 
\begin{equation}
I_i= \frac{2e\bar g_i}{\pi\hbar}\sum_\sigma\integral{\omega}{-\infty}{\infty}
[f_s(\omega)-f_d(\omega)][-{\rm Im}G^r_{d,i,\sigma}(\omega, V_{ds,i})],
\end{equation}
where $\bar g_i=\Gamma_{d,i}\Gamma_{s,i}/(\Gamma_{d,i}+\Gamma_{s,i})$,
$G_{d,i,\sigma}^r(\omega,V_{ds,i})$ is the steady state retarded Green's function
on the dot site, and $f_s(\omega)$, $f_d(\omega)$ are Fermi distribution
functions for the electrons in the source and drain reservoirs, respectively,
$f_{\alpha}(\omega)=f_{\rm F}(\omega-\mu_{\alpha})$ and  $\mu_{s,i}=\alpha_{s,i}eV_i$,
$\mu_{d,i}=-\alpha_{d,i}eV_i$, so that for a difference in chemical potential
across dot $i$ of $eV_i$  due to the bias voltage, $V_i$,
$\alpha_{s,i}+\alpha_{d,i}=1$.

\subsection{Results at $T=0$}
\begin{figure}[!htbp]
\includegraphics[width=0.70\linewidth]{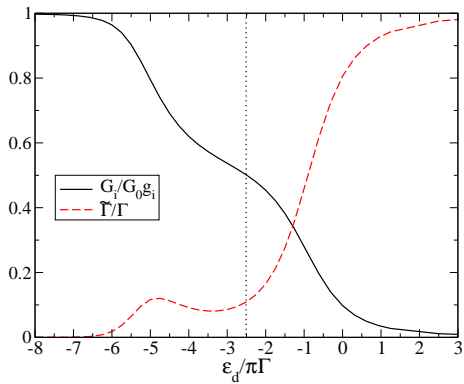}
\caption
{(Color online) 
A plot of $G_i/g_iG_0$ and $\tilde\Gamma/\Gamma$ as a function of $\varepsilon_d/\pi\Gamma$
for the model with $U/\pi\Gamma=20$, $U_{12}/\pi\Gamma=5$ and
  $\pi\Gamma=0.01$. The vertical dotted line corresponds to
  $\varepsilon_d=-U_{12}/2$.}
\label{econd1}
\end{figure}

In the limit of zero temperature and in the absence of a magnetic field, the
zero bias differential conductance through dot $i$, $G_i=dI_i/dV$ reads,
\begin{equation}
G_i= 4\pi \bar g_i\rho_{i}(0)G_0,
\end{equation}
where $G_0={2e^2}/{h}$ is the twice the quantum conductance result.
This can be expressed in terms of renormalized parameters, via
$\rho_{i}(0)=z_i\tilde\rho_{i}(0)$, 
\begin{equation}
G_i= \frac{ g_i G_0}{1+\big(\frac{\tilde\epsilon_{d,i}}{\tilde\Gamma_i}\big)^2},
\label{current}
\end{equation}
where $g_i=4\bar g_i/(\Gamma_{d,i}+\Gamma_{s,i})$. In the spin Kondo regime,
$\tilde\epsilon_{d,i}/\tilde\Gamma_i\to 0$ such that  $G_i\to g_i G_0$ which is
the unitary limit for symmetric coupling to the leads,
$\Gamma_{d,i}=\Gamma_{s,i}$ so  $g_i=1$. At the degeneracy point we have 
$\tilde\epsilon_{d,i}/\tilde\Gamma_i=1$, such that in this case $G_i=g_iG_0/2$.
Generally, in most experimental situations one has $\Gamma_{d,i}\neq
\Gamma_{s,i}$. 
The crossover of the described behavior can be seen in
Fig.~\ref{econd1} where we plot  $G_i/g_iG_0$ as a function of
$\varepsilon_d/\pi\Gamma$ for the parameters $U/\pi\Gamma=20$,
$U_{12}/\pi\Gamma=5$ and $\pi\Gamma=0.01$.

\noindent 
Also plotted is the ratio $\tilde\Gamma/\Gamma$, because this gives a measure
of the width of the quasiparticle resonance to be seen in the spectral density
at zero temperature in terms of that for the non-interacting system $\Gamma$. The quasiparticle
resonance is seen as a peak in the measurement of the  differential
conductance  versus source drain voltage $V$. Our calculations therefore
predict a minimum of the width of the source drain signal when the gate
voltage is tuned along the ridge with enhanced conductance. The conductance
signals in Figs.~2(a,b) of Ref.~\onlinecite{KAWMRKSZG13pre} seem
to indicate the possibility of such a behavior, however, a closer inspection
of the experimental data would be desirable. 
If this resonance is very narrow, as it can be in the spin Kondo regime due to the exponential
renormalization due to $U$, the peak will not be detectable if the resolution of
the temperature of the experiment is such that $T>\tilde\Gamma$. This
is the case in the reported experiments but the peak in the spin/pseudospin degeneracy
regime is seen where the value of $\tilde\Gamma$ is significantly less renormalized than in
the spin Kondo regime (see Sec.~\ref{sec:finT}).

\begin{figure}[!htbp]
\includegraphics[width=0.70\linewidth]{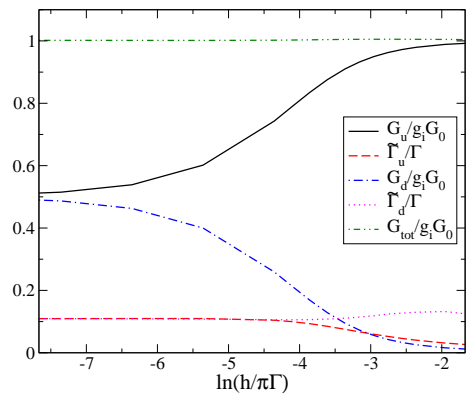}
 \caption
{(Color online) 
A plot the linear conductance of the up and down electrons, $G_{\rm u}/g_iG_0$ and
$G_{\rm d}/g_iG_0$ and and their sum $G_{\rm tot}/g_iG_0$, as a function of ${\rm ln}(h/\pi\Gamma)$
together with the renormalized resonance widths, $\tilde\Gamma_{\rm u}/\Gamma$
and  $\tilde\Gamma_{\rm d}/\Gamma$, for the results shown in Fig. \ref{wilson_ratio_h2}. }
\label{condh}
\end{figure}

If a magnetic field is applied to suppress the spin excitations  in large
fields  the conductance should correspond to the  SU(2) Kondo regime for the
pseudospins. In Fig.~\ref{condh}, we plot the linear conductance in the 
individual spin channels, $G_{\rm u}$ (spin up) and $G_{\rm d}$ (spin down), and the total
$G_{\rm tot}$, as a function of applied magnetic field (log scale) using the
results shown in Fig.~\ref{wilson_ratio_h2}. 

\noindent
The conductance in zero field is that at the degeneracy point where
$G_{\rm u}/g_i G_0=G_{\rm d}/g_iG_0=0.5$, and as the magnetic field is increased
conductance due to the down excitations is suppressed and that due to the up electrons increased
and approaches that for the SU(2) Kondo model. Hence, in this large magnetic
field case we observe spin polarized conductance through the dots which can reach
the unitary limit. The renormalized resonance widths of the up and down
electrons, $\tilde\Gamma_{\rm u}$, and $\tilde\Gamma_{\rm d}$, are shown in the same
figure, that for the up electrons  narrowing 
significantly with increase of field while that for the down electrons
broadens slightly. In this situation where we have treated the dots as
identical the total conductance is independent of the magnetic field.
This means that any deviation from this result would give
information on the differences between the dots and the couplings to their
respective baths.

\begin{figure}[!htbp]
\includegraphics[width=0.70\linewidth]{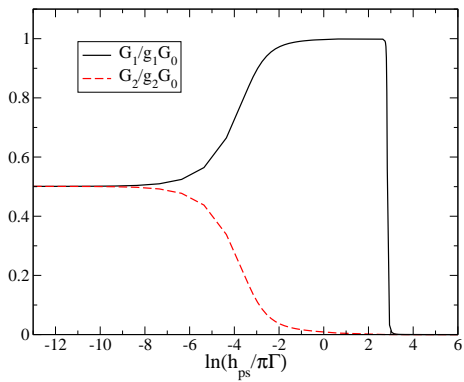}
 \caption
{(Color online) 
A plot the linear conductances $G_1/g_1G_0$ and $G_2/g_2G_0$
 as a function of ${\rm ln}(h_{\rm ps})/\pi\Gamma)$ for the parameters shown
 in Fig.~\ref{wilson_ratio_hps}. } 
\label{condhps}
\end{figure}

In Fig.~\ref{condhps}, we show the conductances of the individual dots
on suppressing the pseudospin excitations by changing the levels on the
individual dots such that $\bar\epsilon_d$ is held constant so as 
to induce a pseudospin field $h_{\rm ps}$. 

\noindent
The results are for the parameter
set given in Fig.~\ref{wilson_ratio_hps}. As the pseudospin field is increased
from the degeneracy point ($\tilde\epsilon_d=\tilde\Gamma$, $n_{d,1}=n_{d,2}=0.5$) there is a
crossover such that the occupation number on dot 1 increases, $n_{d,1}\to 1$,
and that on dots 2 decreases,  $n_{d,2}\to 0$. Over this range the conductance
on dot 1 approaches that of an SU(2) Kondo model due to the remaining
spin degrees of freedom, while that on dot 2 tends to zero. However,
when the pseudospin field reaches values such that both spin states on dot 1
are occupied and  $n_{d,1}\to 2$, the conductance on dot 1 shows a very rapid
crossover such that $G_1\to 0$.\par

\subsection{Results at finite temperature}
\label{sec:finT}
\begin{figure}[!htbp]
\includegraphics[width=0.80\linewidth]{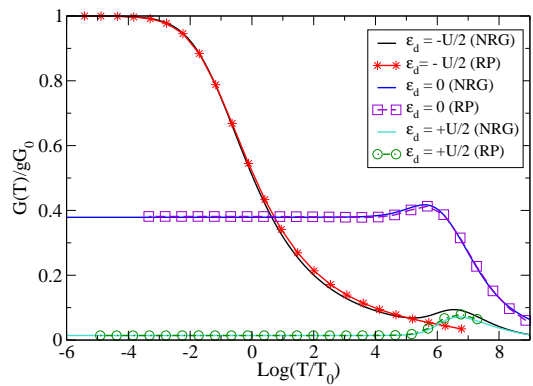}
 \caption
{(Color online) 
The results for linear conductances $G(T)/gG_0$ 
 as a function of ${\rm ln}(T/T_0)$ for the single impurity Anderson model
for $U=16\Gamma$ and $\epsilon_d=-U/2,0,U/2$ given by  Merker et al.\cite{MKMC13} (NRG) compared
with the approximate results based on temperature dependent renormalized
parameters (RP). The value of $T_0$ is defined by $\chi(0)=1/4T_0$, where
$\chi(0)$ is the zero temperature impurity susceptibility. Note that the
particle-hole symmetric case $\epsilon_d=-U/2$  is in the localized
limit where  $T_0=T_{\rm K}$.  }
\label{comp}
\end{figure}

So far we have dealt with the situation at zero temperature. However, the
scale for spin Kondo can be very small such that the finite temperature $T$ in the
experiment matters. A more general expression for the zero bias differential conductance
reads,
\begin{equation}
G_i(T)= \frac{2e\bar
  g_i}{\hbar}\sum_\sigma\integral{\omega}{-\infty}{\infty}\beta\e^{\beta\omega}f_{\rm
  F}(\omega)^2\rho_{d,i,\sigma}(\omega), 
\label{eq:GfinT}
\end{equation}
where $\beta=1/T$. We expect that much of the change of the conductance with
temperature arises from the change in the renormalization of the
quasiparticles on energy scales of the order of the temperature $T$, such
that we can approximate $\rho_{d,i,\sigma}(\omega)$ by the $T=0$ expression but
in terms of  temperature dependent renormalized parameters,
\begin{equation}
\rho_{d,i,\sigma}(\omega)={1\over \pi\Gamma} {\tilde\Gamma^2_{i}(T)\over (\omega-\tilde\epsilon_{d,i}(T))^2+{\tilde\Gamma_i(T)}^2},
\label{rhoT}
\end{equation}
The extension to temperature dependent renormalized parameters was previously
used to calculate the temperature dependence of the spin susceptibility 
for the Anderson model in the Kondo limit and an excellent agreement with 
the exact results from the Bethe ansatz was obtained.\cite{HOM04,Hew05}
The temperature dependence of the parameters can be extracted
from the NRG calculations for an iteration $N$, such that the corresponding
temperature is 
\begin{equation}
  \label{eq:TN}
  T_N = \eta D\Lambda^{\frac{1-N}{2}},
\end{equation}
where $D$ is half the conduction electrons bandwidth and $\eta$ is a constant of order 1. 
We first of all test this approximation for the temperature dependence of
conductance against the NRG results in Fig. 2 in the paper  by Merker et al.\cite{MKMC13}
for the single impurity Anderson model. The results of this comparison  are shown in
Fig.~\ref{comp} for $D=1$, $\Lambda=1.7$ and $\eta=0.55$. 

\noindent
It can be seen that in the most strongly correlated case corresponding
to the particle-hole symmetric model ($T_0=T_{\rm K}$) the agreement is excellent up to
$T=2T_{\rm K}$ and is a  good approximation for $T<150T_{\rm K}$. 
In all three NRG results there is a regime where the conductance
increases with temperature before falling off again at higher temperatures resulting
 in a peak.  The small peak at higher temperatures for the particle-hole
 symmetric case is due to the influence of the atomic peaks at $\omega=\pm
 U/2$, is not seen in the results using quasiparticle approximation for the
 spectral density as the latter does not include these atomic features.
In the less correlated cases away from particle-hole symmetry
the overall agreement is very good and reproduces the peaks seen in the
accurate NRG calculations. The temperature dependence of the renormalized
parameters plays a more important role in the more correlated cases. In the
weakly correlated case $\epsilon_d=U/2$ the temperature dependence of the
parameters plays no role and the peak in the higher temperature regime is due to
the location of the quasiparticle  peak in the spectral density. The position
of the peak for $\epsilon_d=0$ is also due to the location of the
quasiparticle  peak in the spectral density but its height is reduced by the 
temperature dependence of the parameters.  We conclude that the main features
in the  temperature dependence of the differential conductance can be understood in
terms  Eq.~(\ref{eq:GfinT}) using the quasiparticle density of states with
temperature dependent parameters.\par

\begin{figure}[!htbp]
\includegraphics[width=0.90\linewidth]{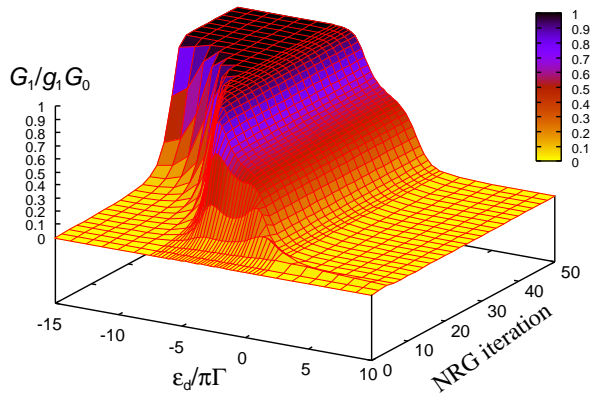}
 \caption
{(Color online) 
A two dimensional plot of the linear conductance ratio $G(T_N)/g_iG_0$ 
 as a function of $\epsilon_d/\pi\Gamma)$ and the NRG iteration number $N$
 which corresponds to a temperature $T_N=0.55\Lambda^{(1-N)/2}$ with
 $\Lambda=6$ for the parameter set $U/\pi\Gamma=20$, $U/\pi\Gamma=5$ 
and $\pi\Gamma=0.01$.
 } 
\label{cond_dd}
\end{figure}

In Fig. \ref{cond_dd} we give a two dimensional plot of the linear conductance
ratio $G(T_N)/g_iG_0$  as a function of $\epsilon_d/\pi\Gamma$ using
renormalized parameters  corresponding to the NRG iteration number $N$ for the
parameter set $U/\pi\Gamma=20$, $U/\pi\Gamma=5$. 

\begin{figure}[!htbp]
\includegraphics[width=0.70\linewidth]{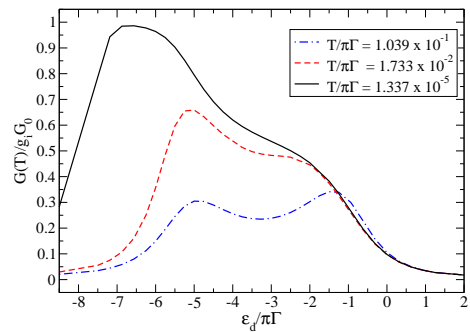}
 \caption
{(Color online) 
The linear conductance $G(T)/g_1G_0$ 
   as a function of $\epsilon_d/\pi\Gamma$
 taken from the results in Fig.~\ref{cond_dd}
for $T/\pi\Gamma=0.1039, 0.01733, 1.337\times 10^{-5}$ corresponding to $N=8,10,18$ } 
\label{cond_dd2}
\end{figure}

\noindent
We estimate the corresponding temperature dependence from the relation in
Eq.~(\ref{eq:TN}), with  $D=1$,  $\Lambda=6$ and $\eta=0.55$. The value of
$\eta$ was selected by the requirement that the calculated entropy $S\to 0$
for large $N$. The effect of increasing temperature (reducing $N$) can be seen to significantly
reduce the conductance in the most strongly correlated regime
$\epsilon_d/\pi\Gamma<-6$ when the temperature exceeds the very small values
of the  Kondo  temperature, and the Kondo resonance in the vicinity of the
Fermi level is suppressed. At higher temperatures ($N\sim 10$) a two peaked response
develops as a function of  $\epsilon_d/\pi\Gamma$. 
This can be seen more clearly in Fig.~\ref{cond_dd2} we extract the results  
for $T/\pi\Gamma=0.1039, 0.01733, 1.337\times 10^{-5}$, which span the
interesting temperature regime. \par 
\bigskip
\noindent
For this parameters set we have estimated $T^*/\pi\Gamma=0.055$ at the
spin/pseudospin degeneracy point, so for $T\sim  T^*$ this falls within the
two peak regime. As the temperature is increased in  the temperature range
$T\sim T^*$, the heights of both peaks are reduced but the height of the peak
at the larger value of $|\epsilon_d|$  decreases  more rapidly. As a
consequence the two peak  structure becomes more symmetrical. At the higher
temperature  $T/\pi\Gamma=0.1039$, where $T\sim 2T^*$, the height of the peak
corresponding to the lower value of $|\epsilon_d|$ begins to become the larger
of the two. 

Though we have not used the particular parameter set for the recently
reported experimental results by Keller et al.\cite{KAWMRKSZG13pre} we find a
two peak form and general trend with temperature  as shown  
in Fig.~2 of their paper (note that the results there are plotted as a
function of $-\epsilon_d$ there). In  Fig.~\ref{cond_dd2} it can be seen that
there is a range $0>\epsilon_d/\pi\Gamma>-1$ where the conductance increases
with  temperature rather than decreases. The behavior is similar in
the results shown in Fig.~2d of Ref.~\onlinecite{KAWMRKSZG13pre}. We conclude
that the quasiparticle picture  with temperature dependent parameters can
provide an explanation of the main features seen in the experimental results.

In Fig.~3 of Ref.~\onlinecite{KAWMRKSZG13pre} also the temperature scaling is
analyzed and NRG calculations for experimental parameters show a small bump at
around 30mK. The peak of the spectral function in this regime is at
$\tilde\epsilon_d$, shifted from the Fermi level $\omega=0$. Therefore, as
also seen in Fig.~\ref{comp} it is possible that the conductance increases at
finite temperature since   
additional spectral weight can become available for transport. A similar effect
effect was observed when the Kondo resonance splits in a magnetic
field.\cite{HBK06} 
It is possible that such an effect is responsible for the 
experimental observation in Ref.~\onlinecite{KAWMRKSZG13pre}. These features,
however, are likely to depend on the particular parameter set used in the
calculations.

\section{Conclusions}
We have surveyed the low energy behavior for a double quantum dot
system described by an Anderson model paying particular attention to the
parameter regime where the spin and 
pseudospin (interdot) excitations become degenerate. In an earlier
theoretical study it has been  asserted that the strong correlation  behavior in this
regime would correspond to that of an SU(4) Kondo model.\cite{BZHHD03} 
To examine this assertion we have calculated the parameters that
specify the effective Hamiltonian for 
the low energy regime, which correspond to renormalized  versions of the
parameters, $\epsilon_{d,i}$, $\Gamma_{i}$, $U_{i}$ and  $U_{12}$,
which describe the original `bare' model. They  can be  accurately
deduced from an analysis of the  low energy excitations of an NRG
calculation.\cite{HOM04,NCH12b} The low energy effective model describes a Fermi liquid 
in which  the quasiparticles interact via the terms,  $\tilde U$ and  $\tilde
U_{12}$. There is a point of 4-fold degeneracy 
for the effective Hamiltonian when the interaction terms between the quasiparticles set to
zero,  $\tilde U=\tilde U_{12}=0$.  For universality and 
 an SU(4) fixed point, however, we require that the low energy response
functions can be expressed in terms of a {\em single} renormalized energy
scale, the Kondo temperature $T_{\rm K}$. Once the interaction terms are
included  the SU(4) symmetry survives only if $\tilde
U=\tilde U_{12}$. For $U\le D$ we find this to be the case only 
if $ U= U_{12}$  so no new symmetry emerges on low energy scale.
This implies that for  $ U>U_{12}$ and $U\le D$
we require two renormalized parameters  to specify the low energy behavior.
For $U\gg D$, there is a regime where we do find SU(4) symmetry
with $U_{12}<U$ provided $U_{12}$ is also greater than, or comparable with, the 
half bandwidth $D$. This is consistent with the derivation of an
SU(4) Coqblin-Schrieffer model based on a Schrieffer-Wolff
transformation.\cite{hewson,MGL06,NCH10a}

We note that there is not a unique SU(4) Kondo model for the double quantum dot.
The Anderson model with  $U_{12}=U$ can  be mapped into
a SU(4) Kondo model also in the case with particle-hole symmetry with
$n_{d,1}=n_{d,2}=1$.\cite{NCH10a} In this case the  operators in the
model correspond to a 6-dimensional representation of SU(4)
 in contrast to the mapping for the spin/pseudospin degenerate model  with
 $n_{d,1}=n_{d,2}=0.5$ where the operators correspond to 
the  fundamental (4-dimensional)  representation of SU(4). \par

The regime with spin/pseudospin degeneracy has attracted experimental
interest\cite{HHWK08,AKRCKSOG13,KAWMRKSZG13pre} as it raises
the possibility of using the pseudospin excitations, which can be manipulated
and observed in independent channels, as a more convenient way to examine
behavior of excitations  in  individual spin species. There are also recent
proposals to use double dot systems for thermoelectric applications\cite{DAH13pre} and create
spin polarized currents\cite{VBAFM13pre} (cf. Fig.\ref{condh}).  
Experimental measurements have been made of electron transport
through the individual dots subject to bias voltages applied to the separate 
conduction electron baths. The results for the conductance as a
function of the bias voltage  correspond to non-equilibrium
steady state conditions and  present a major challenge to theory, because,
though theoretical techniques have been developed successfully to deal
with equilibrium conditions, it has proved to be  difficult to  generalize
them to non-equilibrium situations. The linear response, however, can be
deduced from equilibrium calculations. At $T=0$ the linear response
depends only on the free quasiparticles, and at the degeneracy point the
result does correspond to that for an SU(4) model, which via the Friedel sum
rule can be expressed in terms of the equilibrium occupation numbers on the
dots. However, finite temperature corrections involve the
quasiparticle interaction terms $\tilde U_{i},\tilde 
U_{12}$ and thus will show deviations from universal SU(4) Kondo behavior.
The temperature
corrections to order $T^2$ for the single impurity Anderson model have been
calculated exactly in terms of the renormalized parameters using the
renormalized perturbation expansion (RPT).\cite{Hew93,Hew01,HBK06} Similar
calculations have been carried out for the leading corrections to the linear
voltage regime in powers of  the bias voltage $V$, using RPT in the Keldysh
formulation.\cite{Ogu02,Ogu05,HBO05} The approach should be applicable to the
double dot model but the calculations are lengthy and will be the subject for future work.

We have calculated earlier the leading $T^2$ corrections to the
self-energy for the SU(4) particle-hole symmetric model\cite{NCH12a,NCH12b}
which arise purely from the imaginary part of the self-energy.  
Le Hur et al.\cite{LSL07} have carried out a similar calculation for the non
particle-hole symmetric SU(4) model and find a contribution in this case of the same
order to the real part of the self-energy, such that the temperature corrections
of order $T^2$ cancel out in the expression for the conductance so the leading 
contribution in this case is of order $T^3$. This reflects the fact that the  two SU(4) models
describe different physical situations and as a consequence their leading order
Fermi liquid corrections can differ.

Over the broader temperature scale we have shown that we can estimate the
temperature dependence of the linear conductance based on a spectral density
deduced from the quasiparticle density of states with temperature dependent
parameters. This approach not only predicts features in line with experimental
observations, but also provides a framework for their interpretation.

\bigskip
\noindent{\bf Acknowledgments}\par
\bigskip
\noindent
We would like to acknowledge helpful discussions with T. Costi,
D. Goldhaber-Gordon, B. Halperin, A. Keller, D. Logan and G. Zar\'and.
We thank T. Costi and co-authors for supplying the numerical results on the
temperature dependence of conductance for the Anderson model  from
their paper\cite{MKMC13} for inclusion in Fig. \ref{comp}.  JB acknowledges financial
support from the DFG through BA 4371/1-1. Numerical computation in this work was partially carried out at the Yukawa Institute Computer Facility. This work has
been supported in part by the EPSRC Mathematics Platform grant EP/1019111/1.

\bigskip
\appendix*
\begin{appendix}
\section{Effective Hamiltonians}
An effective Hamiltonian for the model with $U\to\infty$ and
$U_{12}>|\epsilon_{d}|>>\pi\Gamma$ can be found by projecting the full
Hamiltonian onto its atomic (i.e. $\Gamma=0$) groundstates and including the
effects of fluctuations between these groundstates perturbatively to lowest
order in $\Gamma$.  For $V_{\vk,1}=V_{\vk,2}=V_{\vk}$, the impurity
contribution to the resulting effective Hamiltonian is 
\begin{align}
&H_{\rm eff}=\sum_{\vk\vk^{\prime}i}J_{\rm spin}^{\vk\vk^{\prime}}{\bf
  s}_{\vk\vk^{\prime}i}\cdot {\bf S}_{i} \nonumber \\ 
+&\sum_{\vk\vk^{\prime}\sigma\sigma^{\prime}}J_{\perp}^{\vk\vk^{\prime}}
\{(l_{\sigma\sigma^{\prime}}^{\vk\vk^{\prime}})^{+}L_{\sigma^{\prime}\sigma}^{-}+(l_{\sigma\sigma^{\prime}}^{\vk\vk^{\prime}})^{-}L_{\sigma^{\prime}\sigma}^{+}\}
+J_{\parallel}^{\vk\vk^{\prime}}(l_{\sigma\sigma}^{\vk\vk^{\prime}})^{z}L_{\sigma^{\prime}\sigma^{\prime}}^{z}
\label{Heffquarterfilling2}  
\end{align}
where
$S_i^{\alpha}=\frac12\elcre{d,i}{{\sigma}}\sigma^{(\alpha)}_{{\sigma}{\sigma}'}\elann{d,i}{{\sigma}'}$ and
we have introduced the pseudospin raising operator
$L_{\sigma\sigma^{\prime}}^{+}\equiv\ket{1\sigma}\bra{2\sigma^{\prime}}$ and
lowering operator
$L_{\sigma\sigma^{\prime}}^{-}\equiv\ket{2\sigma}\bra{1\sigma^{\prime}}$ and
similarly for the conduction electrons, with
$(l_{\sigma\sigma^{\prime}}^{\vk\vk^{\prime}})^{+}=c^{\dagger}_{\vk,1,\sigma}c^{}_{\vk^{\prime},2,\sigma^{\prime}}$
etc., where $c_{\vk,i,\sigma}=c_{\vk,i,s,\sigma}+ c_{\vk,i,d,\sigma}$
(i.e. appropriate to the situation close to equilibrium).  Here,
$\ket{i\sigma}$ denotes the impurity configuration with one electron of spin
$\sigma=\uparrow,\downarrow$ on the dot $i=1,2$ and the last term in
Eq.\eqref{Heffquarterfilling2} describes a normal Kondo spin exchange
occurring independently on dots $1$ and $2$.  The pseudospin contribution is
anisotropic, with $J_{\perp}^{\vk\vk^{\prime}}=
-V^{}_{\vk}V^{*}_{\vk^{\prime}}U_{12}/\epsilon_{d}(\epsilon_{d}+U_{12})$ and
$J_{\parallel}^{\vk\vk^{\prime}}=2V^{}_{\vk}V^{*}_{\vk^{\prime}}/(\epsilon_{d}+U_{12})$,
whereas the spin contribution is isotropic with an antiferromagnetic exchange
coupling
$J_{\rm spin}^{\vk\vk^{\prime}}=-V^{}_{\vk}V^{*}_{\vk^{\prime}}/\epsilon_{d}$.   
Poor man's scaling equations\cite{And70,BZHHD03,LS03,LSB04} for this effective
model show the mutual influence of spin and pseudospin Kondo physics as visible
in the  renormalization of the respective couplings. 
\end{appendix}

\bibliography{artikel,biblio1}

\end{document}